# Modeling the EdNet Dataset with Logistic Regression

**Philip I. Pavlik Jr. and Luke G. Eglington**

Institute for Intelligent Systems and Psychology, University of Memphis, Memphis, TN, USA
ppavlik@memphis.edu; lgglngtn@memphis.edu

**Abstract**

Many of these challenges are won by neural network models created by full-time artificial intelligence scientists. Due to this origin, they have a black-box character that makes their use and application less clear to learning scientists. We describe our experience with competition from the perspective of educational data mining, a field founded in the learning sciences and connected with roots in psychology and statistics. We describe our efforts from the perspectives of learning scientists and the challenges to our methods, some real and some imagined. We also discuss some basic results in the Kaggle system and our thoughts on how those results may have been improved. Finally, we describe how learner model predictions are used to make pedagogical decisions for students. Their practical use entails a) model predictions and b) a decision rule (based on the predictions). We point out how increased model accuracy can be of limited practical utility, especially when paired with simple decision rules and argue instead for the need to further investigate optimal decision rules.

## Introduction

The EdNet Kaggle competition provides a unique opportunity to determine the best predictive models in the student modeling community. It was very productive for our team to participate because the data's challenges resulted in a great deal of technical skills learning by our team. While we found it a stimulating experience and our results and experiences may be interesting to the community, we also have some feedback about the nature of the task and how this interfaced with our attempt.

## Nature of the Task

The EdNet Kaggle competition's shared task was delivered in such a way that details about the data that would typically be present were missing. The clearest case this was problematic was that descriptions of the items and tags were missing. This lack makes efforts such as ours to create an artificially intelligent model to track learning much more difficult than if we could have used this knowledge to guide our creation efforts. However, whatever the reason for this decision, it seems entirely plausible to assume that it was less troubling for AAAI members with experience in deep knowledge tracing and transformer networks like SAINT and SAINT+ (Choi et al. 2020; Shin et al. 2020). In these other methods and network models, generally, the approach to creating the AI model of the student is made with an approach that itself might be described as AI (rather than our hand-tooled methods founded in statistics like ours). This point is not to claim that we think we would have placed that much better in the competition if we had had such knowledge, but the possibility remains. It was unclear why the information about the data was impoverished enough to restrict a learning science hypothesis-driven feature discovery approach.

The lack of this information may have implications for understanding the models created by the winners of the competition. Ultimately, one might hope that a learner model of students that would function in a learning system would have some type of comprehensible summary statistics that can be presented to students and teachers. This sort of open model is one primary motivation for research in this area (Mitrovic and Martin 2007).

## Tools Used

To accomplish our submissions, we used LKT (Logistic Knowledge Tracing), which is an R package that we (both authors) have been working on and developed further to compete in the shared task (Pavlik Jr and Eglington; Pavlik

Jr et al. 2021, preprint). It is available on GitHub and has some help files that explain how it is used (Pavlik Jr and Eglington). The LKT tool allows the expression of nearly any sort of logistic regression-based model of learning. The tool works as an R function where the user can specify which features of the data correspond to KCs (including items as the most fine-grained sort of KC), and then functions of the history of these features can be used as regression predictors. The simplest example would be the AFM (additive factors model), which captures each performance as a logistic function of the effect of the student (an intercept), KC (an intercept and a coefficient to track the effect of learning for each repetition of the KC).

As it turned out, we needed to significantly enhance this tool to enable the data preparation and models we attempted. At first, the data was difficult to load in R, but this was solved by the datatable package, which is a high-efficiency replacement for R dataframe. We were frankly amazed at the performance improvements in data storage and retrieval and the datatable syntax for executing fast functions on a subset of the data (which was essential in computing derived features).

We also found it necessary to avoid using the R glm package because it was both too slow for models it could compute and incapable of computing any reasonably large model without errors in the model matrix creation step. To accomplish this part of the work, we needed to rebuild LKT to use a sparse model matrix to conserve memory. We also needed to use the LibLineaR package in R, which wraps functions written in C/C++, which accomplish the logistic regression far faster than the R glm command.

## Data Preparation

We realized a bit late how to prepare our data reasonably to simulate how the test set data may have been selected. We realized this process was necessary because our typical modeling methods select full subject histories in a training set and then test with a data set of full histories for different students (student-stratification). This procedure has been adopted because model application in learning technologies has never attempted to update the actual model coefficients as data accumulates. However, this is theoretically possible and might result in a model that behaved more like a deep knowledge tracing model. We did not have this capability in our model.

Because students were in both train and test set for the competition, we could not use our strategy of just using the models built using complete user records since the competition test set had a different selection method than the train set (it was not student stratification; instead it was temporal strata). This validation method meant that models fit on the train data might fit poorly to the test data. We needed a way to pass the histories of practice (so crucial to our method) into the testing environment. This inconvenience of model representation extends to all history features since our methods assume that all history is preprocessed to tabulate the prior amounts/times for each item or skill (often called knowledge components or KCs). This configuration required us to create massive tables for each KC or each student, built from the training data, import these into Kaggle, and then update them based on each batch.

More specifically, our data preparation developed to have the following steps to build a model that would allow for this. We believe the steps are useful to detail because we did not realize R could handle massive datasets before creating this process. We first used the datatable fread command to load in the entire train.csv table. We created a random value between 0 and a year for each user and added that to each user's times. This process simulated the training data being gathered over about one year. At this point, we supposed (or saw in the discussion posts) that the test data came chronologically after the train data; however, it was unlikely that the test data represented the final portion of practice for all users. This ordering was important to consider when deciding how to choose our model training set from the sorted train information in a way similar to how the test data may have been generated. We decided that about 90% through the sequence would be a good rule of thumb, but we did not test that assumption. So, we selected rows 90,000,000 to 92,000,000 from the chronologically sorted train.csv file with the simulated times. This dataset became the simulated test set of similar size to the final Kaggle test. At that point, we found the users who are in this set and pull all the data before that time segment for these users (about 12 million records) to act as a training set if we wanted to see the effect of using features from the training set in our model of the test set. Our use of cumulative features makes it hard to use the Kaggle setup.

Using this setup, we felt that any model we could produce in this simulated test set should closely replicate the Kaggle competition test. Unfortunately, time limitations made it challenging to explore the best way to use this setup fully. One way this may have worked is to give a slightly more accurate application of the intercepts (see below) by allowing us to scale the effect of the training set intercepts (found for the 12M) when applied to predict the test set. This estimating of the correct regression intercepts for generalization seemed conceptually very similar to the concept of mixedeffect regression shrinkage (Quan et al. 2013).

## Key Model Features

Most readers will be familiar with logistic regression, which converts a linear equation into a probability representing a binary event. As mentioned below, item and student ids were highly useful predictors, but it seems that student id and the overall history of success and failure were highly redundant since our attempts to combine them rarely result in much additional gain compared to simply using the intercepts. The small additional gain for more features seemed to be typical of the features we attempted after already having included the content/item id and the student id or overall performance.

We did see that various ways of using the skill tags did improve the model, but they were difficult to use, and the improvements were not large. The simplest way of using the tags was to treat all unique combinations of the tags as KC and simple create a PFA model; however, this was never seen as practical since it would have meant passing a history table from the training set that was too massive for memory (more than 400,000 by 1000). We did see that it may have been possible to accomplish this with sparse matrix packages in R, but the advantage we saw for the effort was consistently small (~AUC improvement of .005-.01). As part of another project, we have a renewed interest in improving domain models for logistic regression, and we attempted to cluster the tags and relabel the KCs based on computing how unique tag combinations covaried (Pavlik Jr. et al. 2008) and then using this information to do fuzzy clustering. The number of clusters then becomes the number of KCs, and each tag is associated with the presence or absence of these KCs. We found a solution with 12 KCs that produced improved fits, but again the improvement was relatively small, and it was also somewhat multicollinear with the unique tag combinations by themselves. We tried many schemes to use the features described here (Pavlik Jr et al. 2021, preprint).

# Results

## Results on Kaggle

Our first models worked remarkably well and gave us a false sense of accomplishment. The first model found a fixed intercept to represent content ids and used 2 coefficients to capture the student performance trend as the function of the overall success and the overall failures for that student. It did not use the prior history of the students in any dynamic way. This model achieved a public score of AUC .741, which we were amazed to see was so high, but it appeared item difficulty and student ability are huge factors as in many educational datasets. Remarkably this model with only the most simple history did the same (AUC .740) as our model with content id intercepts and the student ids (which was a 1-parameter item-response theory model), despite this 2nd model not being adaptive. Unfortunately, combining the two models required fitting a model to the simulated test set where intercepts fit from the training set had been merged in as new features. While we accomplished this as described in the data preparation stage, our competition entries did not test models prepared in this way.

## Results Locally

Without the constraints of the Kaggle submission process, we were able to fit a logistic regression model with the more intricate features that we would typically include in our learner models. Since we knew we had run out of time, we created this model without any student features using our standard procedures. While we did not have time to crossvalidate, since our model does not contain student features, and because it was created with a fit to 10,000 users randomly selected from the train.csv set, it should generalize well.

We included intercepts for content_id and part (part ranging from 1-7 in the dataset) as might be expected. We also included intercepts for each tag combination nested within each part. Intercepts were only estimated for combinations that occurred more than 10 times. Cumulative counts of practice were used as predictors for the number of prior lectures and tag clusters (nested within part). Similarly to the SAINT+ model, we used elapsed time since previous interactions as a predictor (referred to as "recency"). We operationalized recency as a power ($t^{-d}$) decay function of time since the previous interaction. The optimal parameters were found via maximum likelihood estimation. We used recency as a predictor at the level of the tag combination (when did they practice this combination in this part?) and at the level of the clusters (when did the student last practice an item from the present clusters?). The optimal parameter was lower (faster decay) for the cluster-level recency feature. We also included a cumulative count of practice for tag combinations (nested within student) that was recency weighted. We found that including this and the normal count improved fit and that the recency parameter was low, indicating recent trials were weighted more heavily.

Finally, we included a feature that used the prior discrepancy between predictions and outcomes, which we refer to as errordec. This feature is an exponentially decaying average of prior signed errors *by the model*. For instance, if the average signed difference between prior predictions and outcomes is positive, the adjustment by error negative will be negative because positive signed error indicates reliable over-estimation of student performance. The feature is intended to correct for systematic errors induced by student differences in prior knowledge or aptitude. The optimal parameter value for errordec was close to 1 (indicating the signed errors were consistent and not local to recent trials).

This model ended up achieving an AUC of .775. Many other potential features were not explored simply due to time constraints, such as average spacing and weighting elapsed time differently for within vs. between sessions, which may be valuable to investigate in the future.

## Discussion

We conclude with a discussion of two issues in consideration of learner models.

### Prediction accuracy vs. Decision rules

Pedagogical Decision Rules (PDRs) are an important aspect of adaptive learning systems. PDRs describe the criteria by which an item or concept is determined to be worth practicing or not. Common heuristics for this include "Mastery Criterion" (drop item from practice once p(correct)>.95), drop N (practice item unless answered correctly N times in a row), or optimal efficiency thresholds (practice whichever item is closest to the difficulty level that provides optimal learning gains). Learner models are only important to the extent that the PDR requires or uses the model's precision in a meaningful way. However, PDRs are less often the target of research, but they should be because sometimes the choice of PDR can substantially impact learning (Eglington and Pavlik Jr 2020).

The choice of PDR also impacts the relevance of model accuracy. Put another way, if the PDR is sufficiently crude, the models' accuracy will not be as important. For instance, if the PDR is "mastery", the models used will be relevant to the extent to which they differ *when they consider mastery to have been achieved.* In the simplest case, we want to accurately predict what is >.95 probability and should NOT be practiced, and what is below .95 and should be. The learner model's role ends up being to help decide what should and should not be practiced, conditioned on the assumption that the mastery criterion is the correct PDR. As an illustration, we fit two models with the same underlying structure, one with the optimal parameters and one with slightly suboptimal parameters to an educational dataset we frequently use. One has AUC=.775, the other AUC=.763. Both models agree on most of their predictions; they agree on what is >.95 and what is <.95 98.3% of the time. In other words, their input to the PDR will result in the same decisions 98.3% of the time. Why does this matter? It matters because it shows that even if one model has a higher AUC (the difference between the Kaggle winner and 34[th] place), they may differ only very slightly *in practice*. This example is intended to illustrate that a better model according to a prediction metric may be only trivially different in terms of actual practical outcomes (practice what is <.95, ignore what is >.95). Practical outcomes may instead be more driven by the PDR.

More granular PDRs may lead to minor differences in AUC being more relevant to practice selections. PDRs can indeed make significant differences in learning, and Eglington and Pavlik (2020) showed that for vocabulary learning, low difficulty practice selections could result in markedly better learning (but greater difficulty may be preferable for other materials, see (Pavlik Jr. et al. 2020). Eglington and Pavlik showed that practicing whatever item was closest to .86 probability resulted in superior learning to practicing what was closest to .40. This result was primarily due to low probability items being time-consuming due to failure and subsequent feedback being necessary for failures but not successes. As a result, practicing easier items and only introducing harder items when easier items were above the threshold resulted in far more practice trials. The differential time cost for failures and successes is relevant in the EdNet dataset as well: failure trials took 16% longer than success trials, and prior explanations benefited learning while being differentially time-consuming depending on whether the previous trial was correct. Together, these factors point towards correctness probability being relevant for how *efficient* a trial is and suggests the PDR should be chosen with time costs in mind.

### Choice of Decision Rule may Constrain Learner Model Design

In Eglington and Pavlik (2020), practicing according to a target difficulty required using the model to compute *every* item's correctness probabilities after *every* trial. Notably, this regular updating was not possible in the Kaggle challenge (the correctness of previously predicted trials was not always available due to the batched delivery of trials), but such information may be important in many learning contexts in which the outcome of the immediately preceding trial is highly informative (Galyardt and Goldin 2015; Molenaar et al. 2019). Further, there is also a speed issue: *if* the model is to be used to make pedagogical decisions in real-time, it must be fast enough to make predictions about the probability of *all* potential items each trial, if the task is to pick the item closest to a target difficulty.

Trial-by-trial updating for the Eglington and Pavlik model in an adaptive learning system is possible due to logistic regression's computational efficiency and having the model run locally on the user's computer. The model was updated every trial based on the user's practice history. This practical issue of model updating and new predictions for each item is worth considering for more complex models and situations where there may be hundreds of items to make predictions for. How will the model be used in a running system when a running system will likely involve predicting some indicator for some large set of items that might be chosen for the student?

It may be interesting to compare the more complex models that achieved high accuracy in this competition by using different PDRs to adaptively schedule practice. This comparison could allow evaluation versus simpler models to assess the new complex models' practical benefits. We also believe that comparison across different PDRs will highlight just how important they can be. These experiments or simulations will also raise new interesting issues, such as how much data the new model needs to be properly parameterized and how exactly that data should be collected/sampled to facilitate generalization. It is important to see how these alternative approaches to learner models (like SAINT+) account for these issues, especially if it is desirable to have the model generalize to new applied educational contexts. It certainly seems plausible to suppose that if a PDR could be tailored to the unique knowledge encoded in a transformer network, it might boost learning by allowing dynamic practice task selection to optimize the learning effect of complex interactions among skills over time.

**LKT Features for Other Models**

Although other researchers in this competition may prefer deep learning approaches to learner model development, the LKT framework may help researchers without any learning science background understand what types of features have been found to track student learning. Many of the features generated with the LKT package are theoretically informed and may serve as excellent inputs into deep learning models.